\documentclass[a4,aps,pre,twocolumn,10pt]{revtex4-1}
\usepackage[latin1]{inputenc}
\usepackage{amsmath}
\usepackage{amsfonts}
\usepackage{amssymb}
\usepackage{graphicx,hyperref}
\usepackage{amssymb}
\usepackage{mathptmx}

\newcommand{\be}{\begin {equation}}
\newcommand{\ee}{\end {equation}}
\newcommand{\beqa}{\begin {eqnarray}}
\newcommand{\eeqa}{\end {eqnarray}}

\hypersetup{
    colorlinks,    citecolor=black,    filecolor=black,    linkcolor=black,    urlcolor=black
}

\begin{document}

\title{Proton acceleration by circularly polarized traveling electromagnetic
wave}
\author{Amol Holkundkar}
\email{amol.holkundkar@physics.umu.se}
\author{Gert Brodin}
\email{gert.brodin@physics.umu.se}
\author{Mattias Marklund}
\email{mattias.marklund@physics.umu.se}

\begin{abstract}
The acceleration of charged particles, producing collimated mono-energetic
beams, over short distances holds the promise to offer new tools in medicine
and diagnostics. Here, we consider a possible mechanism for accelerating
protons to high energies by using a phase-modulated circularly polarized
electromagnetic wave propagating along a constant magnetic field. It is
observed that a plane wave with dimensionless amplitude of 0.1 is capable to
accelerate a 1 KeV proton to 386 MeV under optimum conditions. Finally we
discuss possible limitations of the acceleration scheme.
\end{abstract}

\maketitle

\affiliation{Department of Physics, Ume{\aa} University, Ume{\aa}, SE-90187,
Sweden}

\affiliation{Department of Physics, Ume{\aa} University, Ume{\aa}, SE-90187,
Sweden}

\affiliation{Department of Physics, Ume{\aa} University, Ume{\aa}, SE-90187,
Sweden}

\section{Introduction}

Laser induced particle acceleration has drawn considerable interest among
researchers all over the world since the pioneering work by Tajima and
Dawson \cite{tajima}. The acceleration gradient of conventional linear
accelerators is of the order of $10^{5}$ V/cm, however today's state of the
art lasers are capable to produce the acceleration gradient many orders of
what can be achieved using conventional LINACs. In general laser based
accelerators can be divided based on the medium in which the acceleration
takes place, which can be either vacuum or a plasma. The vacuum as a medium
for particle acceleration has some inherent advantages over plasma medium.
The problems like instabilities are absent in vacuum, it is easier to inject
the pre-accelerated particle beam in vacuum as compared to the plasma,
collisions of particles with media causing energy loss and beam spreading is
ruled out, etc. Thus we will focus on particle acceleration in vacuum in
this article.

The relativistic motion of the charge particle in large amplitude
electromagnetic (EM) fields are studied in detail by many authors. The
motion of the charged particle in transverse EM wave and the constant
magnetic field along wave propagation is studied by Roberts and Buchsbaum 
\cite{roberts}, which was further extended analytically and experimentally
by Jorv and Trivelpiece \cite{jorv}. More recently the in-depth Hamiltonian
analysis of the dynamics of charge particle in a circularly polarized
traveling EM wave is been studied by Bourdier and Gond \cite{gond}. Various
different schemes have been proposed for the acceleration of charge particle
in traveling EM wave \cite{feng,feng2,kawata,singh}, some includes the
homogeneous magnetic field however some includes the two counter propagating
EM waves.

In this paper we will consider an alternative method to accelerate protons
in vacuum by circularly polarized electromagnetic waves, where the main new
ingredient is a phase-modulation of the EM wave  Emphasis would be on
understanding the dynamics of proton motion under the proposed scheme. The
next section will briefly describe the proposed scheme followed by the
results and discussion.

\section{Model description}

A circularly polarized traveling wave propagating along the $z$ direction is
considered. The electric and magnetic fields of the wave are given by, 
\begin{equation}
E_{x}=E_{0}\ sin[\omega (t-z/c)+\phi (t-z/c)]  \label{ex}
\end{equation}%
\begin{equation}
E_{y}=-E_{0}\ cos[\omega (t-z/c)+\phi (t-z/c)]  \label{ey}
\end{equation}%
and the magnetic fields are expressed as, 
\begin{equation}
B_{x}=(E_{0}/c)\ cos[\omega (t-z/c)+\phi (t-z/c)]  \label{bx}
\end{equation}%
\begin{equation}
B_{y}=(E_{0}/c)\ sin[\omega (t-z/c)+\phi (t-z/c)]  \label{by}
\end{equation}%
with, $\phi (t-z/c)$ is the phase modulation function which is given by 
\begin{equation}
\phi (t-z/c)=\pi \ sin[\alpha \omega (t-z/c)]  \label{phase}
\end{equation}%
where, $\alpha $ is the so called phase modulation factor which controls the
extent of the modulation. A constant magnetic field is also applied along
the direction of wave propagation given by $B_{z}=b_{0}$.

The electric fields denoted by Eq. \ref{ex} and \ref{ey} can be generated by
introducing an \emph{electro-optic phase modulator}. This is an optical
device in which a element displaying the electro-optic effect is used to
modulate the beam of light. The modulation can be done in phase, frequency,
polarization, and amplitude. The simplest kind of modulator consists of a
crystal, such as Lithium niobate whose, refractive index is a function of
the applied electric field \cite{popek}. An appropriate electric field along
a crystal can be applied in such a way that its refractive index modulates,
which eventually will introduce the phase lag in the laser beam. The phase
modulation will also depend on the length of the crystal and other
parameters. Designing an accurate phase modulator for a specific problem may
be an engineering concern, but for the purpose of this article we will
assume that such a problem can be solved satisfactorily. 

\begin{figure}[t]
\centering 
\includegraphics[totalheight=3.8in,angle=270,trim = 2mm 1in 3.2in 1in, clip]
{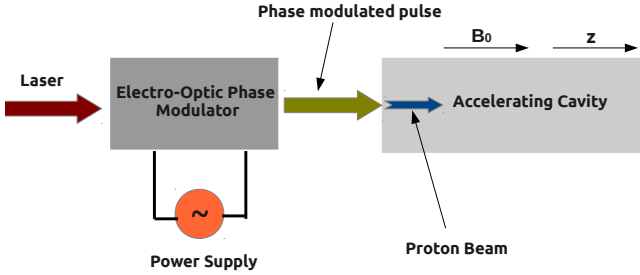}
\caption{Schematic diagram of the proposed scheme.}
\label{geo}
\end{figure}

The schematic diagram for the proposed scheme is shown in Fig. \ref{geo}.
The laser pulse is initially passed through the phase modulator so that the
spatial and temporal dependence of the electric and magnetic fields are
modified according to Eqs. \ref{ex}, \ref{ey}, \ref{bx} and \ref{by}. This
modified pulse is then injected into the accelerating cavity, protons under
the influence of this modified laser pulse undergoes the acceleration along
the transverse direction.

An exact analytical treatment of the problem seems to be too involved
because of the nature of electric and magnetic field profiles. It would be a
non-trivial task to solve the momentum equations corresponding to the field
equations (\ref{ex})-(\ref{by}). In view of this we have numerically
analyzed the dynamics of the particle under the influence of the given field
profiles.

As is wellknown, the motion of the relativistic particle is described by the
following equations,

\begin{eqnarray}
\frac{d\mathbf{p}}{dt} &=&q[\mathbf{E}+\mathbf{v}\times \mathbf{B}] \\
\mathbf{v} &=&\frac{\mathbf{p}/m_{0}}{\sqrt{1+|\mathbf{p}|^{2}/(m_{0}c)^{2}}}
\\
\frac{d\mathbf{r}}{dt} &=&\mathbf{v}
\end{eqnarray}%
where, $\mathbf{p}$, $\mathbf{v}$, $\mathbf{r}$ and $m_{0}$ are relativistic
momentum, velocity, coordinate and mass of the particle. The above equations
are solved numerically by a standard Boris leapfrog scheme where particle
motion is decomposed into motion in the electric field and the motion in the
magnetic field \cite{filippychev}. The particle orbits is calculated by
substituting Eq. (\ref{ex}) -(\ref{by}) into the equation pf motion, and
specifying the initial condition for the injection energy, letting the
initial velocity be directed along $z$. In rest of the paper we have used
the dimensionless units for all physical quantities.

For all the results presented here, the amplitude of the circularly
polarized wave is considered to be $a_{0}=0.1$ (unless otherwise stated),
where $a_{0}=eE/m_{e}\omega c$ \cite{NOTE}. Similarly magnetic field is
denoted by $b_{0}=eB/m_{e}\omega $. Here, $e$ and $m_{e}$ being the charge
and mass of electron, $E$ and $B$ is the amplitude of electric and magnetic
field in SI units with $\omega $ being the laser frequency. The
dimensionless space and time are taken in units of the wave number $k$ and
the angular frequency $\omega $ respectively.

\section{Numerical Analysis}

In this article our main focus is to understand the dynamics of a single
proton with energy 1 KeV (unless otherwise stated), injected along the
propagation \ direction ($z$) of a phase modulated circularly polarized wave
with amplitude $a_{0}=0.1$ and a constant magnetic field with magnitude of $%
b_{0}=1.0$ (Fig. \ref{geo}) along wave propagation. Although it will be
clear later in the article that this scheme is equally valid for the proton
beam having some energy spread. The central theme of the proposed scheme is
the introduction of the so called \emph{phase modulation factor ($\alpha $)}
which can be expressed as the ratio of the phase modulation frequency ($%
\omega _{m}$) to wave frequency $\omega $ i.e. $\alpha =\omega _{m}/\omega $%
. Next we will see how the value of $\alpha $ affects the resulting dynamics
of the particle.

\begin{figure}[b]
\centering \includegraphics[totalheight=3.5in,angle=270]{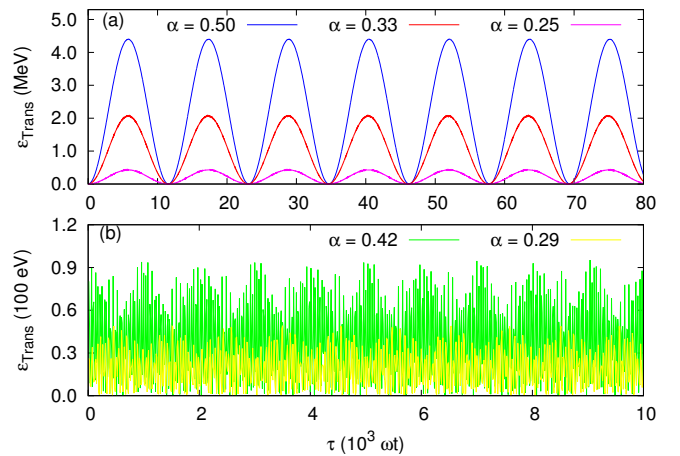}
\caption{Temporal evolution of transverse kinetic energy of proton for (a)
three different phase modulation factor $\protect\alpha = 1/n$ with $n = $
2,3, and 4 being an integer, similarly the case when $n$ is not an integer
is also presented (b). The amplitude of the EM wave $a_0$ = 0.1 and constant
magnetic field along wave propagation $b_0$ = 1.0 is considered.}
\label{transKE}
\end{figure}

The time evolution of the transverse kinetic energy of the particle for the
phase modulation factor $\alpha =1/n$ with $n=2,3$ and $4$ are presented in
Fig. \ref{transKE} (a). It is observed that the proposed scheme of
acceleration works well only when the wave frequency is some harmonics of
the modulation frequency. A large deviation from this condition destroy the
acceleration mechanism completely, Fig. \ref{transKE} (b). Furthermore, as
can be seen from Fig. \ref{transKE}(a), the efficiency of the scheme
deteriorate gradually with higher harmonics of the phase modulation
frequency.

\begin{figure}[t]
\centering \includegraphics[totalheight=3.5in,angle=270]{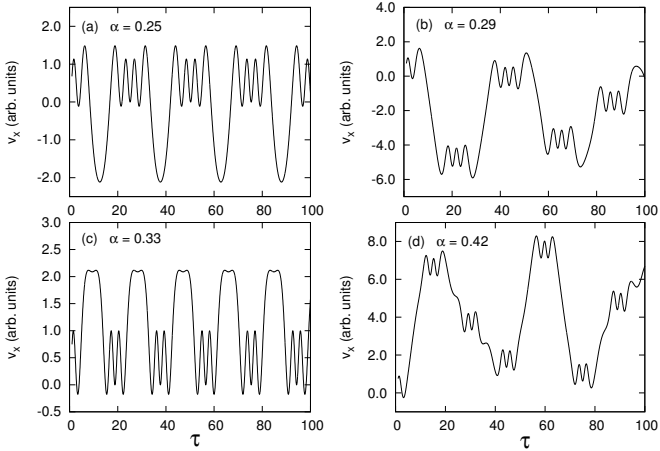}
\caption{Numerical solution of Eq. \protect\ref{modeq} with different values
of $\protect\alpha$ (a) 0.25, (b) 0.29, (c) 0.33 and (d) 0.42 is presented.
considered.}
\label{velo}
\end{figure}

It would be interesting to analyze the underlying principle for the success
of this acceleration scheme. It turns out that the magnetic field as well as
relativistic effects plays only a minor role in this regard. Thus, for the
sake of simplicity let us consider the following equation of motion,

\begin{equation}
m_{i}\frac{dv_{x}}{dt}=q_{i}E_{x}  \label{modeq}
\end{equation}%
where $m_{i}$ and $q$ is the mass and charge of the particle. Only the
temporal dependence in Eq. \ref{modeq} is considered for the purpose of
analysis, which is valid assumption since the dynamics is mostly independent
of the spatial coordinates. Now using the field profile given by Eq. \ref{ex}
and phase modulation function given by Eq. \ref{phase}, this simple equation
of motion (Eq. \ref{modeq})can be integrated to give

\begin{equation}
v_{x}(t)=\frac{q_{i}E_{x}}{m_{i}}\int_{0}^{\tau }sin[\tau ^{\prime }+\pi
sin(\alpha \tau ^{\prime })]d\tau ^{\prime }+C  \label{veloint}
\end{equation}%
where $\tau $ is time in dimensionless units and $C$ is a constant of
integration. For simplification we have chosen $q_{i}E_{x}/m_{i}=1$ and $C=0$%
. The solution of the above integral (Eq. \ref{veloint}) for different
values of $\alpha $ are shown in Fig. \ref{velo}. It is quite clear that if $%
\alpha =1/n$, with $n$ being an integer the temporal evolution of velocity
is very much regular and periodic (Fig. \ref{velo} a,c), however for the
values when $n$ is not an integer the solution of the integral is irregular,
(Fig. \ref{velo} b,d). The observation is that the success and failure of
the proposed acceleration scheme mostly depend on the property of the
integral given by Eq. \ref{veloint}. It is observed that acceleration works
only when the frequency of the electromagnetic wave is some harmonics of the
phase modulation frequency, i.e. $\alpha =1/n$ with $n$ being an integer.

\begin{figure}[b]
\centering \includegraphics[totalheight=3.5in,angle=270]{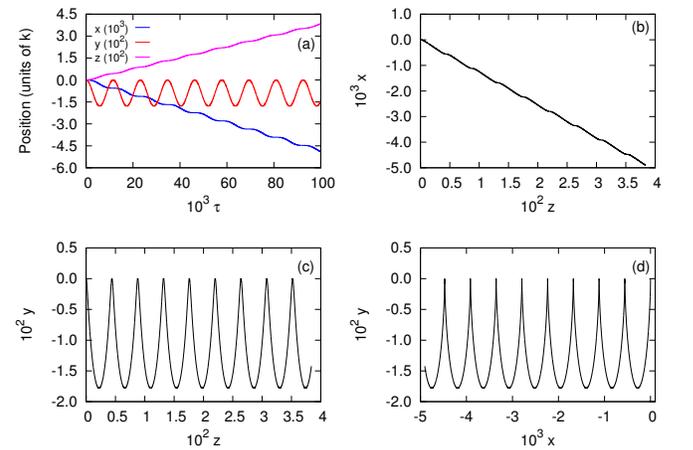}
\caption{ Temporal evolution of the particle trajectory along three space
dimensions (a) is plotted along with the trajectory in x-z plane (b), y-z
plane (c) and x-y plane (d). Here, we have considered the phase modulation
factor $\protect\alpha = 0.5$.}
\label{excurs}
\end{figure}

Now let us focus on the trajectories of the particle under the influence of
the phase modulated EM wave (Eq. \ref{ex}, \ref{ey}, \ref{bx} and \ref{by})
and external magnetic field ($B_{z}=b_{0}$). The time evolution of the
particle trajectory along 3 space dimensions is presented in Fig. \ref%
{excurs} (a). Here we have considered the value of $\alpha =0.5$ (the same
results holds qualitatively for other valid values of $\alpha ,$ i.e. $%
\alpha =0.20,0.25$ and $0.33$).

The longitudinal displacement along $z$ is mainly governed by the energy at
which the particle is being injected in the cavity (magenta curve), however
oscillatory motion along $y$ direction (red curve) is because of the
presence of the external magnetic field along $z$ direction.

As can be infered from the Fig. \ref{excurs} (b) and (d), the biggest
problem with the acceleration scheme is the excursion along the $x$
direction which is much larger than the displacement along the $z$ direction
of propagation. In this scenario the proposed scheme would be impossible to
implement in practice because of the excursion along transverse direction is
too large. In order to deal with this problem a small detuning parameter $%
\delta $ can be added to phase modulation factor $\alpha $, which would be
helpful to confine the particle orbits in the x-y plane. 

\begin{figure}[t]
\centering \includegraphics[totalheight=3.5in,angle=270]{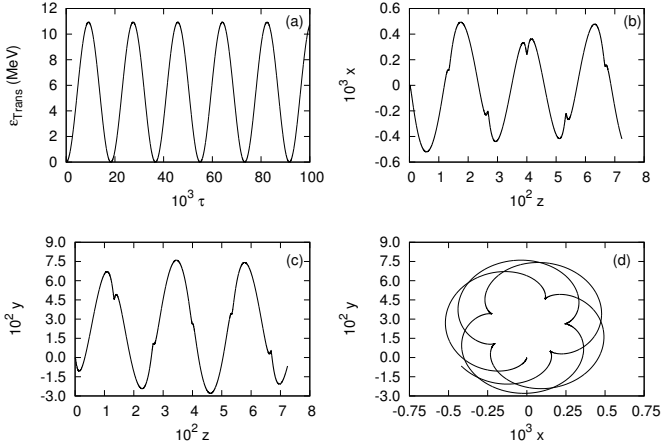}
\caption{ Temporal evolution of the transverse kinetic energy (a) is plotted
along with the trajectory in x-z plane (b), y-z plane (c) and x-y plane (d)
with the detuned phase modulation factor $\protect\alpha = 0.5001$.}
\label{detune}
\end{figure}

Let us examine the trajectory of the particle after the addition of a small
detuning parameter $\delta =10^{-4}$ which modifies the phase modulation
factor $\alpha $ from 0.5000 to 0.5001. It can be observed from Fig. \ref%
{detune}(a) that the kinetic energy of particle increased by a factor of
about 2 as compared to case when no detuning parameter is present (Fig. \ref%
{transKE}). Furthermore the trajectory of the particle is modified
significantly after the introduction of the detuning parameter. As can be
seen, the particle trajectory in the x-y plane (Fig. \ref{detune}(d)) is
closed which make it possible for the acceleration scheme to work in
practice.

It should be noted that the selection of the detuning parameter $\delta $ is
very crucial for the success or failure of the acceleration mechanism. The
value of $\delta $  must be small enough to make the scheme work in favor of
acceleration,  i. e. we must still have the wave frequency to approximately
be a harmonic of the phase modulation frequency, such that the foundation of
the scheme is not destroyed, see Fig. \ref{transKE} (b).  Moreover,  $\delta 
$ cannot be too small, as it must be large enough to prevent the large
excursions (Fig. \ref{excurs} b,d).

\begin{figure}[t]
\centering \includegraphics[width=.7\columnwidth,angle=270]{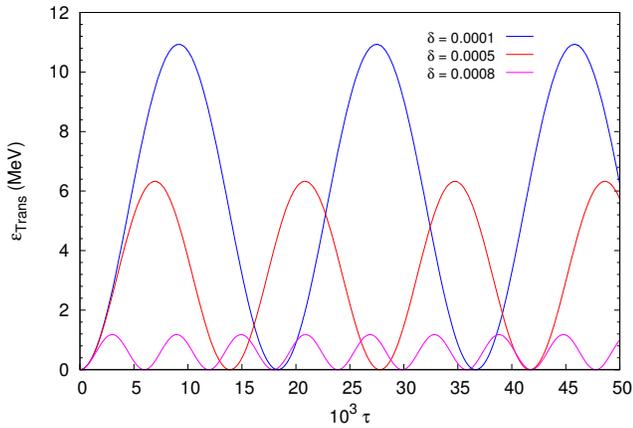}
\caption{ Temporal evolution of the transverse kinetic energy for various
values of detuning parameter $\protect\delta$. The resulting phase
modulation factor would be $\protect\alpha + \protect\delta$ with $\protect%
\alpha = 0.50$.}
\label{ener}
\end{figure}

The effect of the detuning parameter on the energetics of the particle is
presented in Fig. \ref{ener}. As can be seen from this figure, reducing the
detuning parameter from $\delta =0.0008$ to $\delta =0.0001$ increases the
efficiency of the acceleration significantly. On the other hand, the
transverse kinetic energy of the particle is directly related to the
transverse excursion of the particle. For lower values of the $\delta $ the
particle orbits are larger such that the electric field of the wave tend to
do more work for each orbital motion of the particle. As the detuning
increases the particle orbit becomes shorter and shorter, resulting in lower
gain in energy from the wave. Apparently there is an optimum detuning at
which one is able to gain maximum energy per orbital cycle of the particle.

\begin{figure}[t]
\centering \includegraphics[width=.7\columnwidth,angle=270]{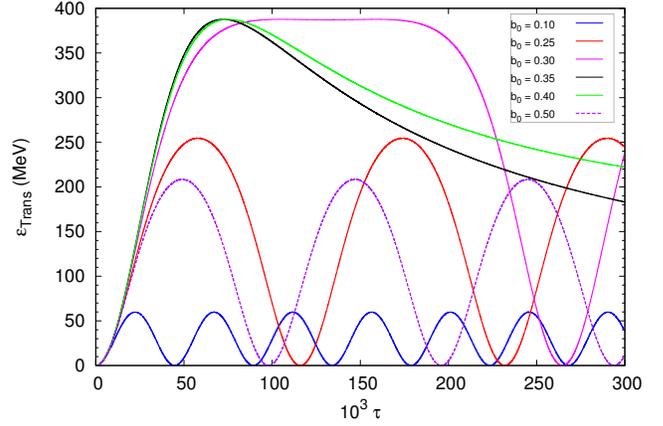}
\caption{ Temporal evolution of the transverse kinetic energy for various
values of applied magnetic fields with detuned phase modulation factor of
0.5001.}
\label{mag}
\end{figure}

\begin{figure}[b]
\centering \includegraphics[width=.7\columnwidth,angle=270]{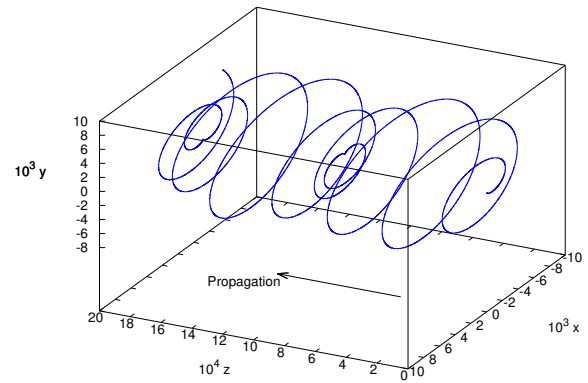}
\caption{Particle trajectory under the influence of applied magnetic field
of magnitude 0.3. The particle is propagating along the $z$ direction.}
\label{orbit}
\end{figure}

So far we have presented all the results with constant magnetic field of
amplitude $b_0 = 1$. It can be understood that the magnetic field is also
responsible along with the detuning parameter to restrict the excursion of
particle in transverse direction. In view of this it would be interesting to
see how the acceleration efficiency varies with the applied constant
magnetic field.

The temporal evolution of the energy varying the constant magnetic field is
presented in Fig. \ref{mag}. The detuned phase modulation parameter in this
case is chosen to be 0.5001. It is observed that the proton can now be
accelerated to energies of about 386 MeV when the applied magnetic field
strength is 0.30. Furthermore, there seems to be an optimum magnetic field
for the maximum acceleration of the particle. This behavior can be explained
on the basis of the Larmor radius of the particle which varies with the
applied magnetic field. The magnetic field should be strong enough to bend
the particle to avoid large excursion and should not be so large such that
the particle orbit is very small and the resulting energy gain in one
orbital motion is small. The particle orbit for the magnetic field $b_{0}=0.3
$ is shown in Fig. \ref{orbit}. As can be seen from Fig. \ref{orbit} the
excursion in the transverse direction is controlled. Thus in this case,
particle acceleration using a phase modulated electromagnetic wave
propagating along a constant magnetic field seems to be possible in a real
scenario.

\begin{figure}[t]
\centering 
\includegraphics[height=3in,width=1.75in,angle=270]{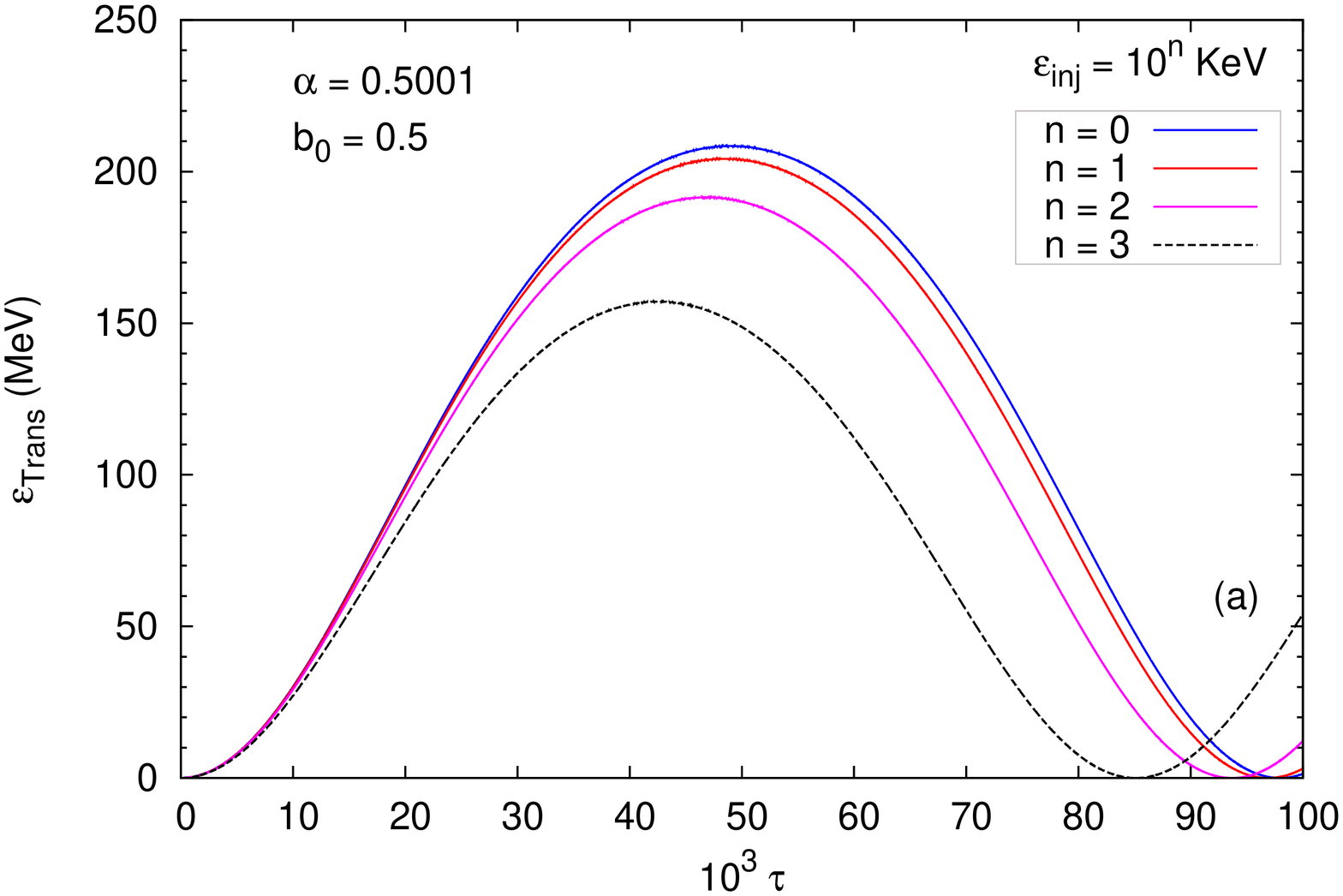} %
\includegraphics[height=3in,width=1.75in,angle=270]{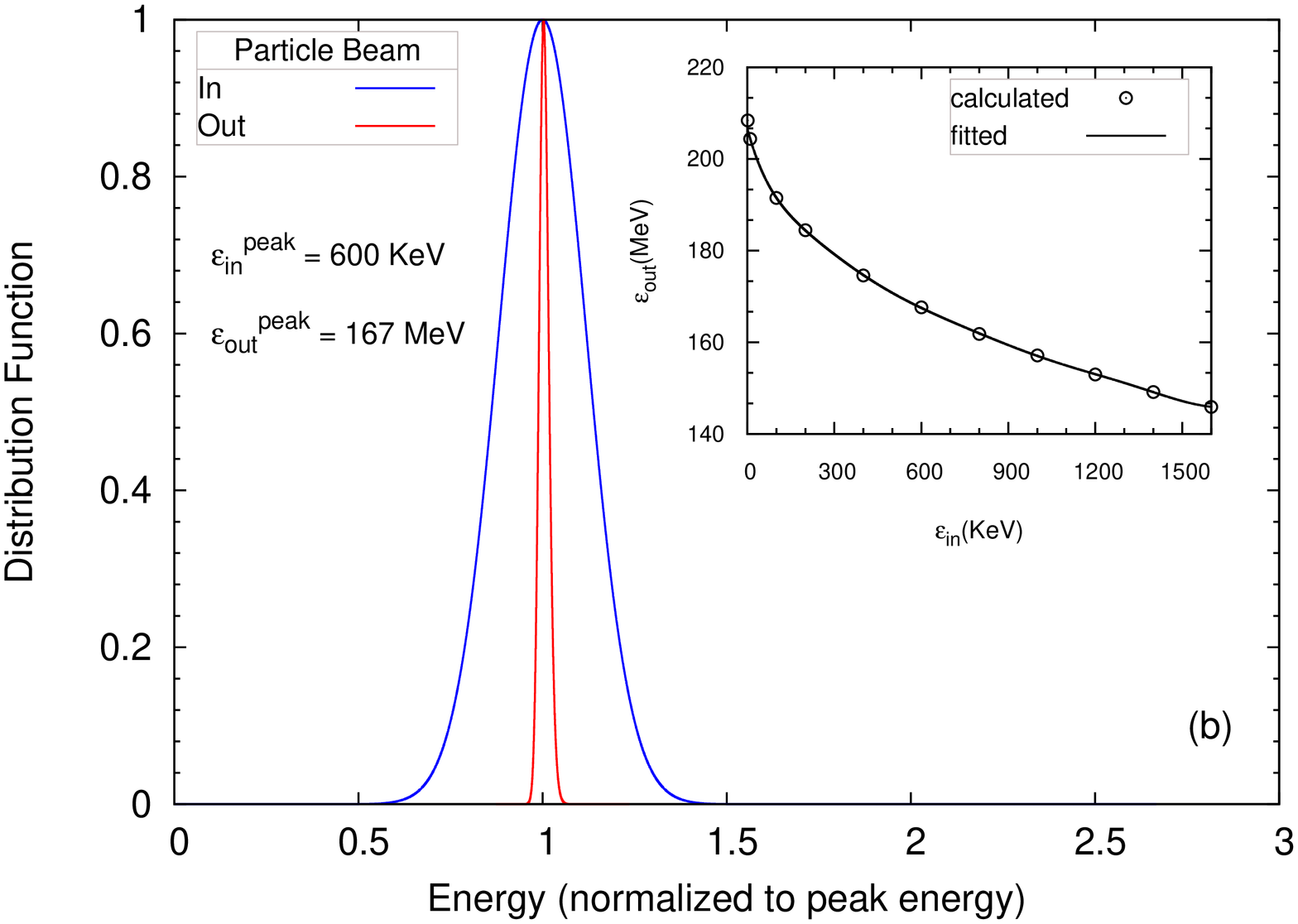}
\caption{Temporal evolution of the kinetic energy for different values of
injection energies (a). A energy spectrum of the output beam is calculated
by injecting a beam with Gaussian energy spectrum (b). The inset in (b)
represents the output beam energy as a function of injection energy, circles
denotes the numerically measured values however solid line is 10$^{th}$
order polynomial fit which is used in calculating the energy distribution
function of output beam.}
\label{inj}
\end{figure}

\section{Discussion}

Here we have focussed on the understanding of the underlying dynamics of the
acceleration. We have established that the phase modulated circularly
polarized wave can accelerate particles in the presence of a constant
magnetic field. There are optimum conditions on the detuning parameter and
the magnitude of the magnetic field, in order to gain maximum energies.
However, for the purpose of applications, the spectral properties plays a
crucial role, in addition to the maximum energy. In view of this it would be
important to understand how this scheme works for a particle beam with an
energy spread, instead of just a single particle. In order to shed some
light on the acceleration dynamics of the particle beam we have separately
simulated the single particle motion when injected with different energies,
which can be perceived as the energy spectral spread of the beam.

The time evolution of the transverse kinetic energy of single particle is
presented in Fig. \ref{inj}(a) for different injection energies. The phase
modulation factor is considered to be $\alpha =0.5001$ and the magnitude 0.5
is considered for the constant magnetic field. It can be observed that even
if the injection energies are varied an order of magnitude, the output
energy of the beam is not very much affected, keeping the scheme functional.
The most important property of the particle beam is its energy distribution
function. The energy distribution function of the output beam is calculated
in Fig. \ref{inj}(b) by considering the input beam having Gaussian energy
spectrum with peak energy of 600 KeV and FWHM of 166 KeV, the energy spread (%
$\Delta E$/$E_{peak}$) of the input beam is about 28\%. The peak energy of
the output beam is observed to be 167 MeV with FWHM of 5 MeV, with the
energy spread about 3\%.

The inset in Fig. \ref{inj}(b) shows the variation of output energy with the
injected particle energy. Circles denotes the actual numerical value which
is fitted with a 10$^{th}$ order polynomial (solid line) in order to find
the energy spectrum. The decline in the output particle energy with input
energy can be explained on the basis of the interaction time of the particle
with the EM wave. The faster the particle, the lesser is the interaction
time with the EM wave, and hence the energy transfer to the particle.

As we have observed in Fig. \ref{inj}(b), the output particle energy is more
or less independent of the injected particle energy, which is apparently
visible in the almost monoenergetic energy distribution of the output beam. 

\section{Final remarks}

A further evaluation of the feasibility of the proposed acceleration scheme
need to go beyond the 1D-variations of the electromagnetic fields. In
particular, it is clear that picking parameter values corresponding to
intense lasers give succesful results for the maximum particle energies.
However, a potential limitation is that large transverse particle excursions
exclude the use of very focused pulses, in which case the available laser
intensity drops accordingly. To some extent this may be remedied with a
strong value of the static magnetic field, which limits the transverse
excursion. In practice, magnetic field stengths well beyond $100$ $\mathrm{T}
$ is needed if the system works in the optical laser regime, which makes
this regime less attractive. Decreasing the wave frequency in the scheme
reduces the need for extreme magnetic field strengths, since we may allow
for somewhat larger particle excursions. The optimal frequency regime may
lie in the infra-red regime or lower, but a full 3D analysis is needed to
optimize the parameters in a realistic scenario. This remains a project for
further research.

\acknowledgments
This work is supported by the Baltic Foundation, the Swedish Research
Council Contract \# 2007-4422 and the European Research Council Contract \#
204059-QPQV. This work is performed under the \emph{Light in Science and
Technology} Strong Research Environment, Ume{\aa} University.

{}


\begin{thebibliography}{99}
\bibitem{NOTE} It should be noted that the $a_0$ is normalized by electron
mass, though the dynamics of the proton will be governed by the field
amplitude which is about 1836 times less than $a_0$, resulting in
non-relativistic energy regime.
\end{thebibliography}
\end{document}